\documentclass[journal]{IEEEtran}
\usepackage{cite}

\ifCLASSINFOpdf 
\usepackage[pdftex]{graphicx}
\else
\usepackage[dvips]{graphicx}
\fi

\usepackage[cmex10]{amsmath}
\usepackage{algorithmic}
\usepackage[ruled,norelsize]{algorithm2e}

\usepackage{array}

\usepackage{mdwmath}
\usepackage{mdwtab}
\usepackage{todonotes}

\usepackage{eqparbox}

\ifCLASSOPTIONcompsoc
\usepackage[caption=false,font=normalsize,labelfont=sf,textfont=sf]{subfig}
\else
\usepackage[caption=false,font=footnotesize]{subfig}
\fi

\usepackage{fixltx2e}
\usepackage{graphicx}
\usepackage{amsfonts}
\usepackage{amssymb}
\usepackage{amsthm}
\usepackage{bm}
\usepackage[ruled]{algorithm2e}
\usepackage[utf8]{inputenc}
\usepackage{booktabs}
\usepackage{epstopdf}
\usepackage{tabularx}
\usepackage{cite}
\usepackage{flushend}
\usepackage[hyphens]{url}

\definecolor{maroon}{cmyk}{0,0.87,0.68,0.32}

\newcounter{tempEquationCounter} 
\newcounter{thisEquationNumber}

\begin{document}
%
\title{Simultaneous Transmit and Receive Operation in Next Generation  IEEE 802.11 WLANs:  A MAC  Protocol Design Approach}

\author{Adnan~Aijaz,~\IEEEmembership{Member,~IEEE},
        and~Parag~Kulkarni
\thanks{The authors are with the Telecommunications Research Laboratory, Toshiba Research Europe Ltd., Bristol, BS1 4ND, UK. Contact e-mail: adnan.aijaz@toshiba-trel.com}}
\markboth{IEEE Wireless Communications -- Accepted for Publication}%
{Shell \MakeLowercase{\textit{et al.}}: Bare Demo of IEEEtran.cls for Journals}
%


\maketitle
\begin{abstract}
\boldmath
Full-duplex (FD) technology is likely to be adopted in various legacy communications standards. The IEEE 802.11ax working group has been considering  a simultaneous transmit and receive (STR) mode for the next generation wireless local area networks (WLANs). Enabling STR mode (FD communication mode) in 802.11 networks creates bi-directional FD (BFD) and uni-directional FD (UFD) links. The key challenge is to integrate STR mode with minimal protocol modifications, while considering the co-existence of FD and legacy half-duplex (HD) stations (STAs) and backwards compatibility. This paper proposes a simple and practical approach to enable STR mode in 802.11 networks with co-existing FD and HD STAs.  The protocol explicitly accounts for the peculiarities of FD environments and backwards compatibility. Key aspects of the proposed solution include FD capability discovery, handshake mechanism for channel access, node selection for UFD transmission,  adaptive acknowledgement (ACK) timeout  for STAs engaged in BFD or UFD transmission, and mitigation of contention unfairness.   Performance evaluation demonstrates the effectiveness of the proposed solution in realizing the gains of FD technology for next generation WLANs.

\end{abstract}



%
\IEEEpeerreviewmaketitle

\section{Introduction}
\IEEEPARstart{R}{ecent} advances in self-interference cancellation \cite{FD_SIC} have led to the practical realization of full-duplex (FD) radios that can transmit and receive simultaneously, unlike conventional half-duplex (HD) radios. Existing efforts towards  FD wireless networks have mainly investigated the Physical (PHY) layer aspects \cite{FD_Phy_Mac}; however,  medium access control (MAC) layer solutions have also started to emerge. To achieve  true benefits of FD wireless communications, enhancements and optimizations are required at different layers of the protocol stack. 

\begin{figure}
\centering
\includegraphics[scale=0.24]{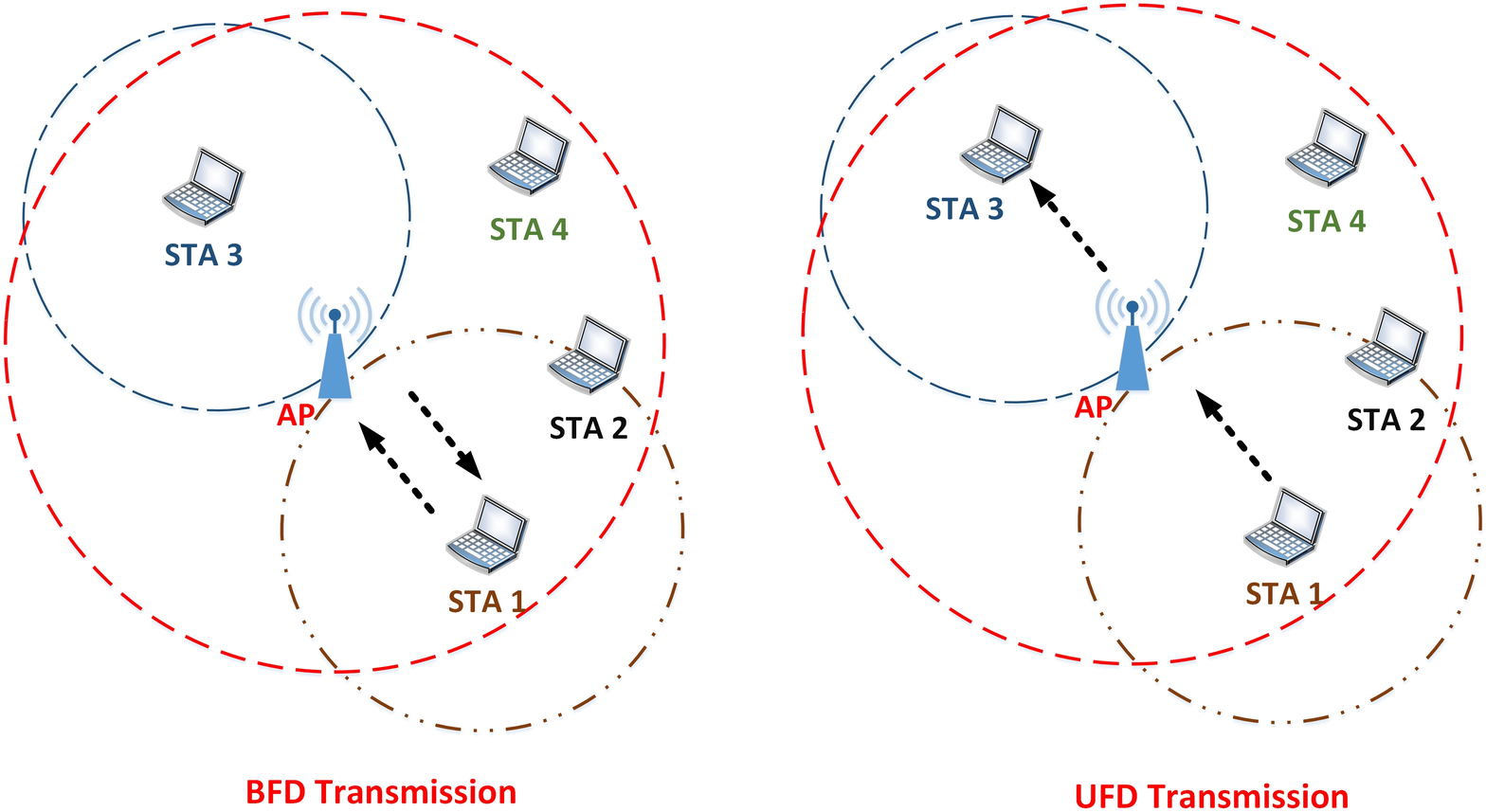}
\caption{An illustration of STR mode in 802.11 networks.}
\label{fd_scenarios}
\end{figure}

On the other hand, IEEE 802.11 wireless local area networks (WLANs) have been quite successful in fulfilling the requirements of low-cost wireless Internet access in public, domestic, and business scenarios. The next generation WLANs face the challenges of providing increased user throughputs and operating successfully in dense deployments. To address these challenges, the IEEE 802.11ax-2019 task group \cite{11ax} is developing new PHY and MAC layer enhancements for improving the performance of WLANs. 

Because of its attractive features, FD technology is  under consideration for adoption in a range of legacy communications standards. In this respect, the IEEE 802.11ax  working group  has been considering a simultaneous transmit and receive (STR) mode (FD communication mode) for the next generation  WLANs \cite{11ax_mag}. By allowing the access point (AP) and the station (STA) to operate in the STR mode, the channel capacity can be theoretically doubled. By enabling STR mode in 802.11 networks, two distinct types of wireless links are created: (a) bi-directional FD (BFD) in which a pair of FD-capable AP and STA can simultaneously transmit/receive to/from each other, (b) unidirectional FD (UFD) in which the AP can simultaneously transmit to a FD/HD STA while receiving from another FD/HD STA. Both types of FD links are illustrated in Fig. \ref{fd_scenarios}.

Enabling STR mode in 802.11 networks creates a number of new challenges. These challenge, which have been discussed in detail later, must be addressed with minimal protocol modifications, while accounting for the co-existence of FD and legacy HD\footnote{The legacy HD mode, throughout this paper, refers to 802.11 Distributed Coordination Function (DCF) \cite{DCF} based WLAN.}  STAs and backwards compatibility. Research efforts for integrating STR mode in 802.11 networks are in infancy. To the best of our knowledge, no concrete solution has been proposed so far that explicitly addresses the design challenges of enabling STR with FD/HD co-existence and backwards compatibility.  \textcolor{black}{Our objective in this paper is to develop a simple and practical solution to enable STR mode in 802.11 networks while accounting for the co-existence of FD and legacy HD STAs, peculiarities of FD environments, and backwards compatibility.} We begin our discussion by covering the state-of-the-art in terms of related MAC protocols. After that we outline the key design challenges which arise by enabling STR mode in 802.11 networks. Centered around these challenges, we propose a novel MAC protocol for enabling STR mode. The proposed protocol particularly accounts for the co-existence of FD and legacy HD STAs, and achieves the benefits of FD communications without affecting backwards compatibility. Finally, we conduct a performance investigation of the proposed protocol.

\section{State-of-the-Art}\label{sect_rw}
This section covers state-of-the-art in terms of MAC protocols for enabling FD communications in infrastructure-based wireless networks. 

In \cite{janus}, Kim \emph{et al.} developed Janus, which is an AP-based MAC protocol for FD wireless networks. Janus uses a centralized mechanism to collect information from STAs and schedule transmissions. Janus supports both BFD and UFD transmissions. However, it does not account for the co-existence of HD and FD nodes. Prototype implementation shows that Janus can achieve nearly \(2.5\) times throughput improvement over HD. 

Sahai \emph{et al.} \cite{SRB_MAC} proposed a FD MAC protocol  based on IEEE 802.11 packet structure with a new FD header and three new protocol elements: shared random back-off, header snooping, and virtual contention resolution. The protocol considers both BFD and UFD transmissions. However, it is not backwards compatible. Prototype implementation shows throughput gains of up to \(70\%\) compared to a HD network. 

In \cite{a-duplex}, Tang and Wang introduced A-Duplex, which is an IEEE 802.11-based MAC protocol  for  co-existence of HD STAs and FD AP.  It exploits packet-alignment based capture effect to establish dual links between the AP and two different STAs.  A-Duplex requires building an interference map of the network, which yields significant overhead. It also introduces a new field in the RTS control frame header of legacy 802.11 DCF, which affects backwards compatibility. 

Duarte \emph{et al.} \cite{fd_mac} designed a MAC protocol,  based on IEEE 802.11 specifications, to support HD and FD nodes. The protocol utilizes request-to-send/clear-to-send (RTS/CTS) messages for FD discovery and opportunistic data transmission. The protocol only supports BFD transmissions. Performance evaluation claims throughput gains of more than \(80\%\) compared to a HD  network. Although the proposed protocol accounts for co-existence of HD and FD nodes, it is not completely backwards compatible because of modifications to ACK management and overhearing behavior for legacy nodes.

In \cite{PoCMAC}, Choi \emph{et al.} developed a power-controlled MAC for FD Wi-Fi networks wherein only the AP operates in FD mode. To overcome the interference between STAs, the authors introduce the concepts of distributed interference measurement distributed power control. The protocol introduces signal strength based backoff mechanism and additional short control frames for coordinating and completing FD transmissions, which affect backwards compatibility.

A MAC protocol proposal \cite{csr_mac} for STR in 802.11ax networks was also presented at one of the TGax meetings. The proposed MAC protocol is designed for the co-existence of FD AP and legacy HD STAs. Unlike 802.11 DCF, the protocol is based on Enhanced Distributed Channel Access (EDCA) mechanism. The STR capability is realized through modifications to legacy procedures which do not ensure complete backwards compatibility. 

\textcolor{black}{Marlali and Gurbuz \cite{scw-fd} proposed synchronized contention window FD (S-CW FD) MAC protocol. The protocol is based on IEEE 802.11 DCF specifications and accounts for the co-existence of FD and HD STAs. However, it  requires additional modifications, like exchange of backoff window size information, which affect backwards compatibility. }



\section{Key Design Challenges for Enabling STR}\label{sect_chal}

 
\subsection{FD/HD Co-existence} 
In practice, not all the nodes in the network will be equipped with FD radios and therefore, FD nodes will co-exist with legacy HD nodes. In order to initiate FD transmissions, FD nodes (APs and STAs) must be able to discover FD capabilities in an autonomous manner.  Such capability discovery must not come at the expense of any modifications to legacy frame structure of 802.11 protocol. Besides, any protocol modifications to support FD/HD co-existence must ensure backwards compatibility. 


\subsection{Enabling BFD and UFD Transmissions}
Enabling STR mode in 802.11 networks results in BFD and UFD transmissions. BFD can theoretically double the link capacity between FD-capable nodes in the network. On the other hand, UFD transmission provides the opportunity of utilizing the FD capability of APs among other HD/FD STAs which provides \textcolor{black}{network-level throughput} improvement. UFD is particularly attractive when the density of FD-capable STAs is relatively lower. In order to reap maximum benefits of the FD technology, both BFD and UFD transmissions must be enabled. However, this must not come at the expense of modifications to legacy channel access protocols affecting backwards compatibility for HD transmissions. 


\subsection{Node Selection for UFD Transmission} \label{elig_node}
After discovering the FD capabilities, any FD-capable STA can  engage in a BFD transmission with the AP and vice versa. However, not all the STAs (either HD or FD) in the network can become part of a UFD transmission. The two nodes simultaneously served by the AP, in UFD case, must be out of the interference range of each other as otherwise the first transmitter (STA A transmitting to the AP in Fig. \ref{fd_scenarios}) will interfere with the receiver (STA B receiving from the AP in Fig. \ref{fd_scenarios}) of the second transmitter (AP in Fig. \ref{fd_scenarios}). Therefore, the AP must know which nodes are eligible to become part of the second transmission in UFD case. Such functionality for the AP must incur minimal overhead.
 
\subsection{ACK Timeout}
Another important challenge for nodes engaged in FD transmissions is to send and receive ACKs for successful transmissions. In legacy 802.11 networks, nodes expect an ACK after sending a data packet. If an ACK is not received during the ACK timeout period, the sender node retransmits the data packet. In case of BFD transmission, data packets are simultaneously sent  by two nodes, and each node gets data packets before getting an ACK. Hence, any of the nodes engaged in BFD transmission, may start waiting for an ACK (after finishing transmission) while the other node may still be engaged in transmission, which leads to an ACK timeout. The issue of ACK timeout becomes particularly challenging in case of UFD transmission as the FD node (AP in this case) cannot simultaneously transmit \emph{or} receive to/from two different nodes. Therefore, the MAC protocol must consider the peculiarities of BFD and UFD transmissions, and implicitly account for the ACK timeout issue without affecting backwards compatibility for legacy HD nodes. Moreover, a single solution is desirable for both BFD and UFD cases.

\subsection{Contention Unfairness}

By enabling BFD and UFD transmissions, contention unfairness issue may arise in the network. Consider that a FD STA (STA 1) is engaged in a BFD transmission with the AP.
While this BFD transmission is going on, a nearby STA (STA 2) will receive erroneous/corrupted packets due to the interference arising from simultaneous reception of packets from STA 1 and AP.  After the completion of BFD transmission, both AP and STA 1 will wait for DCF interframe space (DIFS) duration before next contention. However, STA 2 will wait for extended interframe space (EIFS) duration before next contention, resulting in unfairness in channel access, since EIFS duration is larger than DIFS duration. In legacy 802.11 networks, EIFS is defined (for a STA to defer its channel access following the reception of corrupted packets) to allow extra time for the intended receiver (who may have received the data correctly) to return an ACK without interference. Note that the contention unfairness issue affects both HD and FD STAs and is present in case of UFD transmission as well. Therefore, a key challenge for enabling STR mode in 802.11 networks is to mitigate such contention unfairness by modifying the behavior of overhearing STAs without affecting backwards compatibility.  



\section{MAC Protocol Design -- Key Aspects}\label{sect_protocol}


In order to address the aforementioned challenges, we design a novel MAC protocol. The key aspects of the proposed solution are described as follows. 

\begin{itemize}
\item \textbf{FD Capability Discovery} -- In HD/FD co-existence scenarios, FD nodes (APs and STAs) must be able to discover FD capabilities in an autonomous manner. The proposed solution achieves this through no modifications to the frame structure, but with embedding additional information in existing \emph{reserved} fields of management frames. This approach ensures complete backwards compatibility for legacy nodes. 
\item \textbf{Eligible Node Identification} -- It is particularly important for the AP to know which nodes are eligible to become part of the second transmission in UFD case. The proposed solution enables this functionality through a simple procedure based on neighborhood information. The procedure can be implemented during the network initialization phase. 
\item \textbf{CTS-FD Control Message} -- To initiate BFD and UFD transmissions, the proposed solution utilizes the legacy CTS control message with 1-bit modification in any of the \emph{reserved} bits. Such modification is completely backwards compatible as the legacy HD nodes will be agnostic to the information transmitted in the reserved bits. 
\item \textbf{Adaptive Transmission and ACK Timeout} -- In order to tackle the ACK timeout issue (for nodes engaged in FD transmissions), the proposed solution utilizes a  novel approach for setting the parameters of second transmission and ACK timeout at the nodes engaged in BFD and UFD transmissions. This approach is completely backwards compatible as HD nodes set the ACK timeout in the legacy way. 
\item \textbf{Contention Fairness} -- The contention unfairness issue is actually a by-product of enabling FD communications in the network. The proposed protocol utilizes two simple techniques to mitigate contention unfairness in the network. 
\end{itemize}

\begin{figure}
\centering
\includegraphics[scale=0.4]{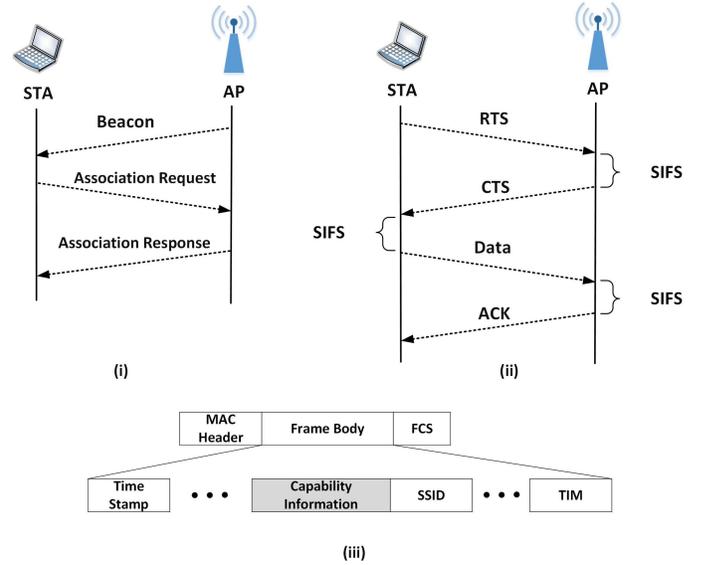}
\caption{Signaling exchange in legacy 802.11 networks: (i) association with the AP; (ii) RTS/CTS  message exchange for data transmission; (iii) capability information field within the general 802.11 management frame.}
\label{legacy_prot}
\end{figure}

\section{MAC Protocol Operation}
In order to describe the proposed MAC design, we consider a single-cell multi-user scenario wherein both FD and legacy HD STAs co-exist.  \textcolor{black}{The legacy MAC protocol operation is  based on IEEE 802.11 DCF \cite{DCF} specifications which employ carrier sense multiple access with collision avoidance (CSMA/CA) and binary exponential backoff algorithms. The signaling exchange in legacy 802.11 DCF WLANs is shown in Fig. \ref{legacy_prot}. Typically, the AP broadcasts the beacon frame (with network information) at periodic intervals. A STA associates with the AP through exchange of association request and response frames.  Before transmitting a data packet, a transmitting node reserves the medium through RTS/CTS handshake with a receiving node. Other nodes in the network, overhearing the RTS/CTS messages, set their network allocation vector (NAV) to the duration specified in RTS/CTS messages and defer access to the medium.  A receiver node,  after receiving the data message, waits for short interframe space (SIFS) duration before sending an ACK. If the sender does not receive an ACK during the ACK timeout period (which is set when sending the data message), it retransmits the data message.} Further, we assume that the FD-capable nodes in the network employ necessary self-interference cancellation techniques at the PHY layer. 

Initially, nodes in the network must be able to discover the FD capabilities. We exploit the \emph{Capability Information} (CI) field (illustrated in Fig. \ref{legacy_prot}) of the management frame  for FD capability discovery by the AP and the STAs. The AP can periodically advertise its FD capability in the beacon frame. Therefore, STAs in the network can learn if the AP is FD-capable or not. The CI field comprises 2 bytes, out of which 1 byte is reserved and is currently used to advertise various capabilities like encryption, QoS, MIMO, etc. The FD capability can be advertised through a 1-bit change in any of the reserved bits. A FD STA can also inform the AP of its FD capabilities when sending the association request frame. The FD capability can be advertised through a 1 bit change in any of the reserved bits of the 2 byte CI field within the association request frame.

%


After discovering the FD capabilities, any FD-capable STA in the network can engage in a BFD transmission with the AP. However, not all the STAs can become part of a UFD transmission. Hence, the AP must know which nodes are eligible to become part of the second transmission in UFD case. We propose a simple procedure for the AP to acquire this knowledge. The key aspect of this procedure, which can be implemented during network initialization phase, is the neighborhood information exchange between STAs and AP. The AP sends an RTS to each STA  in the network. The respective STA (e.g., STA A) responds  with a CTS. Other STAs overhear the CTS and maintain a neighborhood information table. For example, STAs lying within the interference range of STA A overhear the CTS and add the ID of STA A in the neighborhood
information table. At the end of the network initialization phase, each STA reports its neighborhood table to the AP. Based on the overall neighborhood information, the AP learns which STAs are eligible to become part of the UFD transmission. \textcolor{black}{The AP can update the neighborhood information table periodically. This can be achieved in a standards-compatible way through the IEEE 802.11k \cite{802_11k} amendment which supports radio resource measurement functionality in a request-report manner. The AP can trigger a measurement request-response phase at periodic intervals.   }


Next, we explain the procedure for establishment of legacy HD and FD transmissions in our proposed protocol. Consider the system model depicted in Fig. \ref{fd_scenarios} and assume that STA 1 has data to send to the AP. Note that if the AP is HD-capable, only HD transmissions will take place and therefore, the legacy procedure is followed irrespective of the capability of STA 1. If the AP is FD-capable and STA 1 is HD-capable then there is a possibility of either a HD transmission between AP and STA 1 or a UFD transmission involving the AP, STA 1, and another (eligible) STA. If both STA 1 and AP are FD-capable then three possibilities arise: (i) BFD transmission between STA 1 and AP, (ii) UFD transmission involving the AP, STA 1,  and another STA, and (iii) HD transmission between STA 1 and AP  (if AP does not have data to send to STA 1).



In order to initiate BFD and UFD transmissions, we introduce a new type of CTS control message, termed as CTS-FD. Before that, we briefly explain one of the key fields in RTS and CTS (and ACK)  messages. This is the ‘Duration ID’ field
(hereinafter referred to as the ‘duration’ field) that specifies the total transmission time (in \(\mu\)s) required for a frame. For example, the duration in RTS message is set to the time needed to transmit data, CTS, and ACK messages with all the SIFS intervals. The STAs receiving RTS/CTS read the duration field and set their NAV which indicates how long they must defer from accessing the medium. 

The CTS-FD control message differs from the legacy CTS control message by 1 bit i.e., any reserved bit\footnote{The FC field contains many reserved bits e.g., Type value of ‘11’ or Sub-type values from ‘0000-1001’. } in the \emph{Frame Control} (FC) field (which precedes the duration field) of the CTS control message is set to 1 to create the CTS-FD control message. 
Therefore, the CTS-FD control message is completely
backwards compatible with legacy HD STAs and its practical realization is not a challenge.



\begin{figure*}
\centering
\includegraphics[scale=0.35]{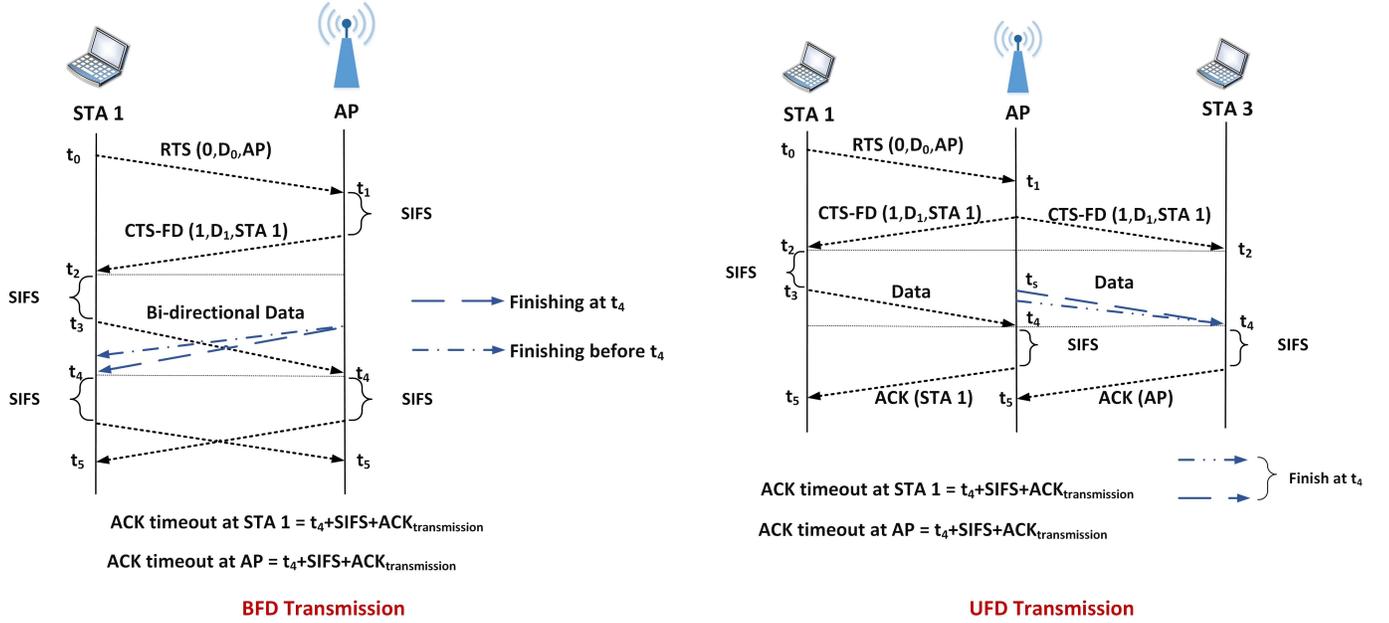}
\caption{BFD (left) and UFD (right) transmission scenarios in the proposed MAC protocol. }
\label{BFD_UFD_txn}
\end{figure*}

\subsection{BFD Transmission}
The BFD transmission scenario, which is illustrated in Fig. \ref{BFD_UFD_txn}, is described as follows. Consider that STA 1 has data to send to the AP. It sends an RTS message to the AP at time \(t_0\). The three fields associated with the RTS message in Fig. \ref{BFD_UFD_txn} correspond to the reserved bit of the FC field (0 indicates legacy), the duration field, and the destination address, respectively. The duration field in RTS is set to \(D_{\text{RTS}}=D_{0}=3\times\text{SIFS}+T_{\text{CTS}}+T_{\text{Data}}+T_{\text{ACK}}\),
where \(T_{\text{CTS}}\), \(T_{\text{Data}}\), and \(T_{\text{ACK}}\) denote the transmission time of the CTS, the data frame, and the ACK, respectively. The RTS message reaches the AP at time \(t_1\). If the AP has data to send to STA 1, it responds with a CTS-FD message after waiting for SIFS duration. The CTS-FD sent by the AP has the reserved bit of FC field set to 1
and includes the destination address of STA 1. The AP uses \(D_0\) as the basis for calculating \(D_1\), which is the duration field in the CTS-FD message. The value of \(D_1\) is calculated as \(D_{\text{CTS}}=D_{1}=D_0-\left(T_{\text{CTS}}+\text{SIFS}\right)\).

The CTS-FD informs STA 1 of a possible FD transmission. After waiting for a SIFS duration, STA 1 starts data transmission. Similarly, after sending the CTS-FD message, the AP waits for SIFS duration and sends a data message (addressed to STA 1). Therefore, a BFD transmission occurs between STA 1 and the AP. Consider that the data transmission from STA 1 to the AP ends at time \(t_4\), which the AP knows based on \(D_0\) or \(D_1\), i.e., \(t_4=t_1+D_0-(\text{SIFS}+T_{\text{ACK}})\). As discussed later, it is particularly important that the data transmission from the AP to STA 1 ends before or at \(t_4\). Therefore, the AP must select the payload and the modulation and coding scheme (MCS) for this transmission accordingly, in order to fulfil this timing constraint as well as the signal-to-noise-ratio (SNR) requirements at the STA. 

Next, we explain the ACK timeout setting procedure and also elaborate on the constraint for suitable selection of the payload and the MCS. Assume that the data transmission from STA 1 to the AP (first transmission) finishes before the data transmission from the AP to STA 1 (second transmission). In this case, the AP cannot acknowledge the transmission of STA 1 as it is already engaged in transmission. Therefore, STA 1 may unnecessarily re-transmit the data owing to an ACK timeout. Similarly, STA 1 cannot acknowledge the second transmission, if it finishes before the first transmission. Therefore, the minimum ACK timeout duration for STA 1 and the AP is equal to \(t_5\), which is the sum of \(t_4\), SIFS, and \(T_\text{{ACK}}\). By setting the ACK timeout to \(t_5\), STA 1 and AP do not need to unnecessarily retransmit. Since the first transmission is initiated by STA 1, the AP must select the function of payload and MCS, which determines the transmission time, that guarantees the second transmission will finish before or with the first transmission (i.e., before or at \(t_4\)).

\subsection{UFD Transmission}
The UFD transmission scenario, illustrated in Fig. \ref{BFD_UFD_txn}, is described as follows. Consider that STA 1 has data to send to the AP.  It sends an RTS message to the AP at time \(t_0\). However, the AP has data to send to STA 2 and STA 3. Therefore, the AP can potentially engage in a UFD transmission. Since STA 2 is within the interference range of STA 1 (see Fig. \ref{fd_scenarios}), it would not be selected by the AP to take part in the UFD transmission.  The AP responds to the RTS by transmitting a CTS-FD message with the destination address of STA 1 and duration field set to \(D_1\). The neighboring STAs, receiving the CTS-FD message, do not know if a BFD or UFD transmission will take place. Note that the STAs receiving the RTS message or the CTS-FD message set their NAV according to duration advertised therein i.e., \(D_0\) or \(D_1\), respectively. Since the CTS-FD message contains the destination address of STA 1, it starts transmitting data at time \(t_3\). The AP knows when this data transmission from STA 1 will end, i.e., at time \(t_4\). Since STA 3 is eligible to take part in the UFD transmission, the AP sends a data message (with the destination address of STA 3) at time \(t_s\). Other neighboring STAs follow their NAV after finding out that the data is intended for STA 3. Therefore, a UFD transmission is successfully established by the AP. As discussed later, it is particularly important that the second transmission (from AP to STA 3) ends at time \(t_4\). Therefore, the AP must select the payload and MCS for this transmission according to the aforementioned time constraint. The AP must adjust the start time of second transmission (\(t_s\)) to meet this constraint. Let, \(t_{est}\) denote the data transmission time to STA 3, which is the function of payload and MCS. If \(t_{est}=t_4-(t_2+\text{SIFS})\), then \(t_s=t_2+\text{SIFS}\). On the other hand, if \(t_{est}<t_4-(t_2+\text{SIFS})\), then \(t_s=t_4-t_{est}\). 

Next, we explain the ACK timeout setting procedure. Assume that the first transmission (STA 1 to AP)  finishes before the second transmission (AP to STA 3).  In this case, the AP cannot acknowledge the transmission of STA 1 as it is already engaged in transmitting to STA 3. Therefore, STA 1 may unnecessarily re-transmit the data owing to an ACK time out. Similarly, assume that the first transmission finishes after the second transmission. In this case, although STA 3 can acknowledge the transmission, it will result in a collision since the AP is already engaged in receiving from STA 1. 
Therefore, the minimum ACK timeout duration for STA 1 and the AP is equal to \(t_5\), which is the sum of \(t_4\), SIFS, and \(T_\text{{ACK}}\).

As per 802.11 specifications, \emph{after a successful reception of a frame requiring
acknowledgment, transmission of the ACK frame shall commence after a SIFS period, without regard to the busy/idle state of the medium}. Therefore, despite its NAV set to expire at \(t_5\), STA 3 will be able to return the ACK after reception of the data from the AP is complete. Note that the proposed mechanism is applicable whether STA 3 is FD-capable or HD-capable. Therefore, any node can become part of the UFD transmission.


\emph{Remark 1} -- The constraint on the AP for adjusting the start time of second transmission to finish at \(t_4\) is only applicable to UFD transmission, and arises because the node selected for second transmission (STA 3 in this case) could be a HD node. As per the current 802.11 DCF specifications, whenever a node receives a data message, it waits for SIFS duration and sends back an ACK. Therefore, a transmission from the AP ending before or after \(t_4\) will result in the aforementioned ACK issues. Hence, through adjusting the start time of second transmission, either HD or FD node can become part of the second transmission, which maximizes the achievable gain of UFD transmission. 


\emph{Remark 2} -- The AP applies some  fairness metric if more than one  STA is eligible to take part in the UFD transmission. 

\textcolor{black}{\emph{Remark 3} -- The selection between BFD and UFD transmissions is an implementation-specific issue. In case an opportunity arises for both, the AP will perform some decision making with respect to a specific criteria  (fairness among users, network-level throughput, etc.); however, this is outside the remit of the proposed protocol. }

\subsection{AP-initiated Communication Scenario}
Consider that the AP in Fig. \ref{fd_scenarios} is FD-capable and has data to send to STA 1. If STA 1 is HD-capable, only HD transmission is possible and the legacy procedure takes place.  If STA 1 is FD-capable, a BFD transmission can take place. The AP sends an RTS message to STA 1. 
If STA 1 has data to send to the AP, it responds with a CTS-FD message which notifies the AP of a BFD transmission. The same procedure is followed for bi-directional data transmission and ACK timeout setting as described earlier for the BFD transmission. If STA 1 has no data to send to the AP, it responds back with a CTS message, and the legacy procedure is followed. Note that the UFD transmission is not possible in case of AP-initiated communication scenario. 


\subsection{Mitigation of Contention Unfairness}
The contention unfairness issue arises as a result of enabling FD communications in the network. We propose two different solutions to mitigate the contention unfairness issue. The first solution provides \emph{a priori} knowledge of a FD-transmission and modifies the overhearing behavior of FD-capable nodes such that, on receiving the CTS-FD  message, they ignore any corrupted packets which would be received during the NAV period.  However, this solution only addresses the contention unfairness issue among FD-capable nodes and is effective when the density of FD nodes is  much higher than the density of HD nodes in the network. The legacy nodes continue to be affected by the contention unfairness issue. In the second solution, the AP sends an FD Transmission Indicator (FDTI) message after completion of a BFD or UFD transmission. Upon receiving an FDTI message, a node which has set the waiting time for next contention to EIFS, as a result of a FD transmission, will cancel the EIFS timer and resume contention after DIFS duration. Note that this solution solves the contention unfairness issue for both legacy and FD-capable nodes. The FDTI message can be  realized through any reserved bits in the MAC header. Further, no change is needed at the PHY layer of legacy nodes; only a software upgrade is required.

\begin{center}
	\begin{table}
		\caption{Key Parameters for Performance Evaluation}
		\begin{center}
			\begin{tabular}{ll}
				\hline	
				\toprule
				Parameter  & Value \\\hline
				\midrule
				Density of APs (\(\lambda_a\)) & \(1.5\times 10^{-4} \ \text{m}^{-2}\) \\
				Density of STAs (\(\lambda_u\)) & \(3\times 10^{-3} \ \text{m}^{-2}\) \\				
				Channel bandwidth & \(20 \ \)MHz \\
				Transmit power & \(40 \ \text{dBm}\) (AP); \(30 \ \text{dBm}\) (STA)   \\
				Traffic model & Backlogged \\
				Header size & \(272\) bits (MAC); \(128\) bits (PHY)  \\
				Payload & \(10\) kbits \\
				Control packet size & \(288\) bits (RTS); \(240\) bits (CTS-FD)  \\
				ACK packet & \(240\) bits \\
				Slot time & \(20\) \(\mu\)s \\
				Interframe duration  &  \(10 \ \mu\)s (SIFS); \(50 \ \mu\)s DIFS   \\
				Transmission rate & \(1 \ \)Mbps (control);   \(54 \ \)Mbps (data)  \\
				RSI parameters \cite{exp_FD} & \(\Delta=38 \ \text{dB}\); \(\chi=13 \ \text{dB}\)  \\ 
				\hline
			\end{tabular}
		\end{center}
			\label{params_numerical}
	\end{table}
\end{center}

\section{Performance Evaluation}\label{sect_perf}
We conduct a performance evaluation of the proposed protocol through system-level simulation studies. Our customized simulator implements IEEE 802.11 DCF specifications and the proposed protocol for enabling STR mode.  We consider Poisson distributed APs and STAs in an area of \(800 \times 800 \ \text{m}^2\), with density \(\lambda_a\) and \(\lambda_u\), respectively. Further, \(\lambda_u=\lambda_f+\lambda_h\), where \(\lambda_f\) and \(\lambda_h\) denote the density of FD-capable and HD-capable STAs, respectively. We assume that users are associated with their closest AP in terms of received power.  The channel model accounts for large-scale path loss and small-scale Rayleigh fading. Other simulation parameters are given in \tablename~\ref{params_numerical}. The transmission range of the AP and the STA is set to \(80 \ \text{m}\) and \(20 \ \text{m}\), respectively. The simulation framework accounts for both protocol model and  physical model. The former captures successful activation of a transmitter node after carrier sensing and contention whereas the latter  considers success of a transmission based on link-level signal-to-interference-plus-noise-ratio (SINR).   We adopt an experimentally characterized model \cite{exp_FD} for residual self-interference (RSI) in FD nodes. 

\begin{figure*}
\label{combined_results}
\centering
\subfloat[]{\label{STR_Gain}\includegraphics[scale=0.25]{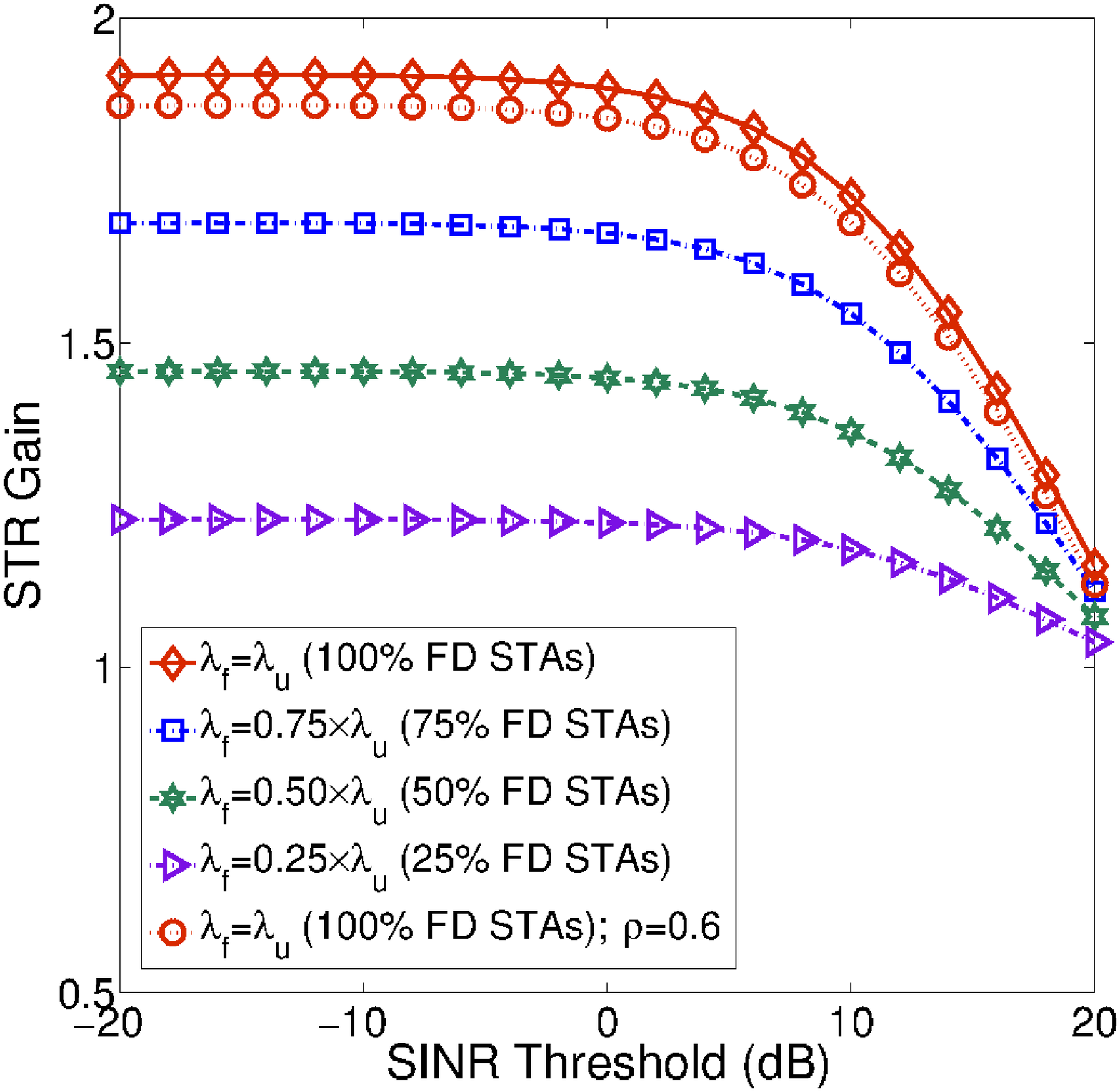}} \qquad
\subfloat[]{\label{STR_Gain_UFD}\includegraphics[scale=0.25]{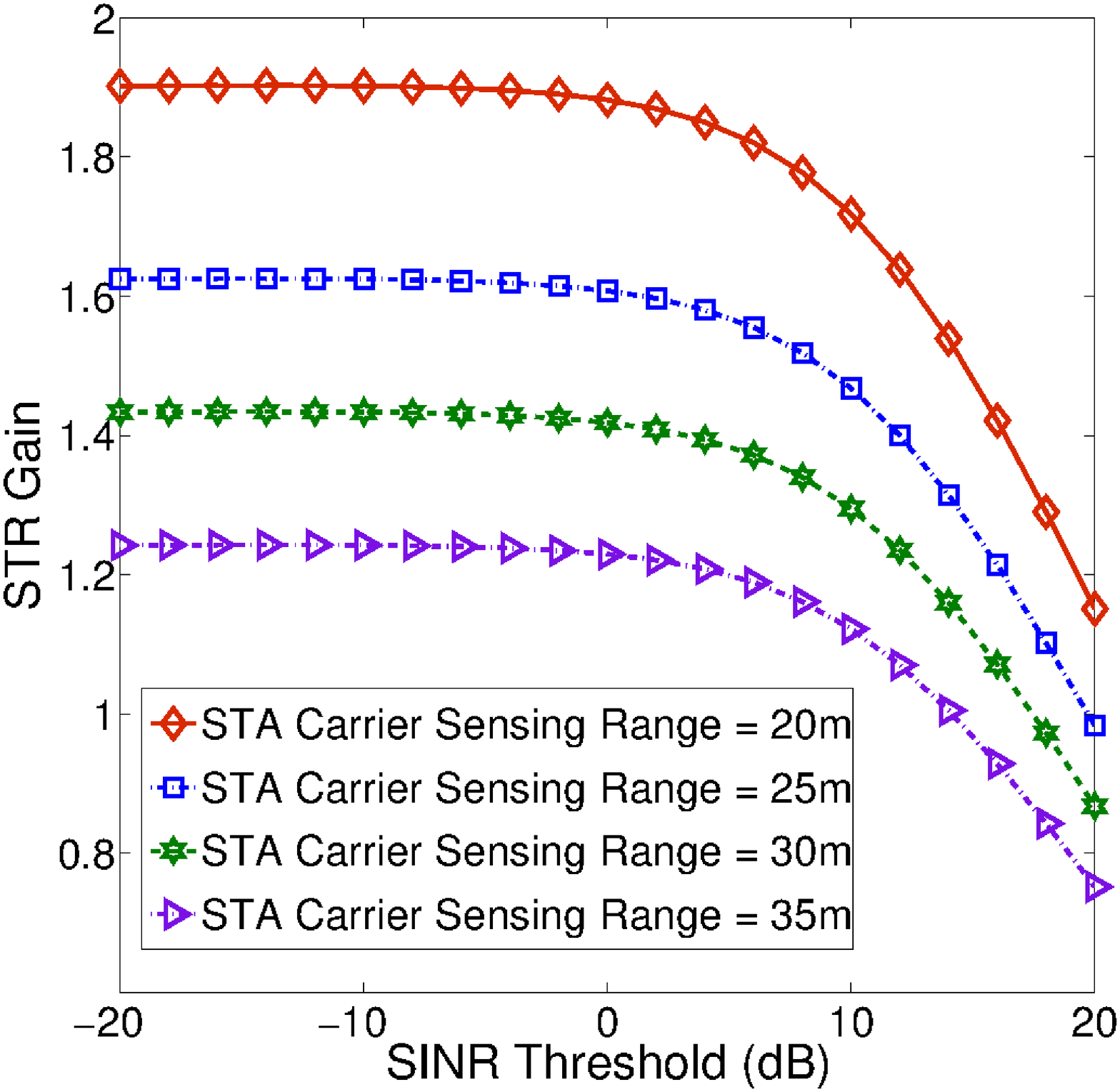}}\qquad
	\subfloat[]{\label{gpt_gain}\includegraphics[scale=0.25]{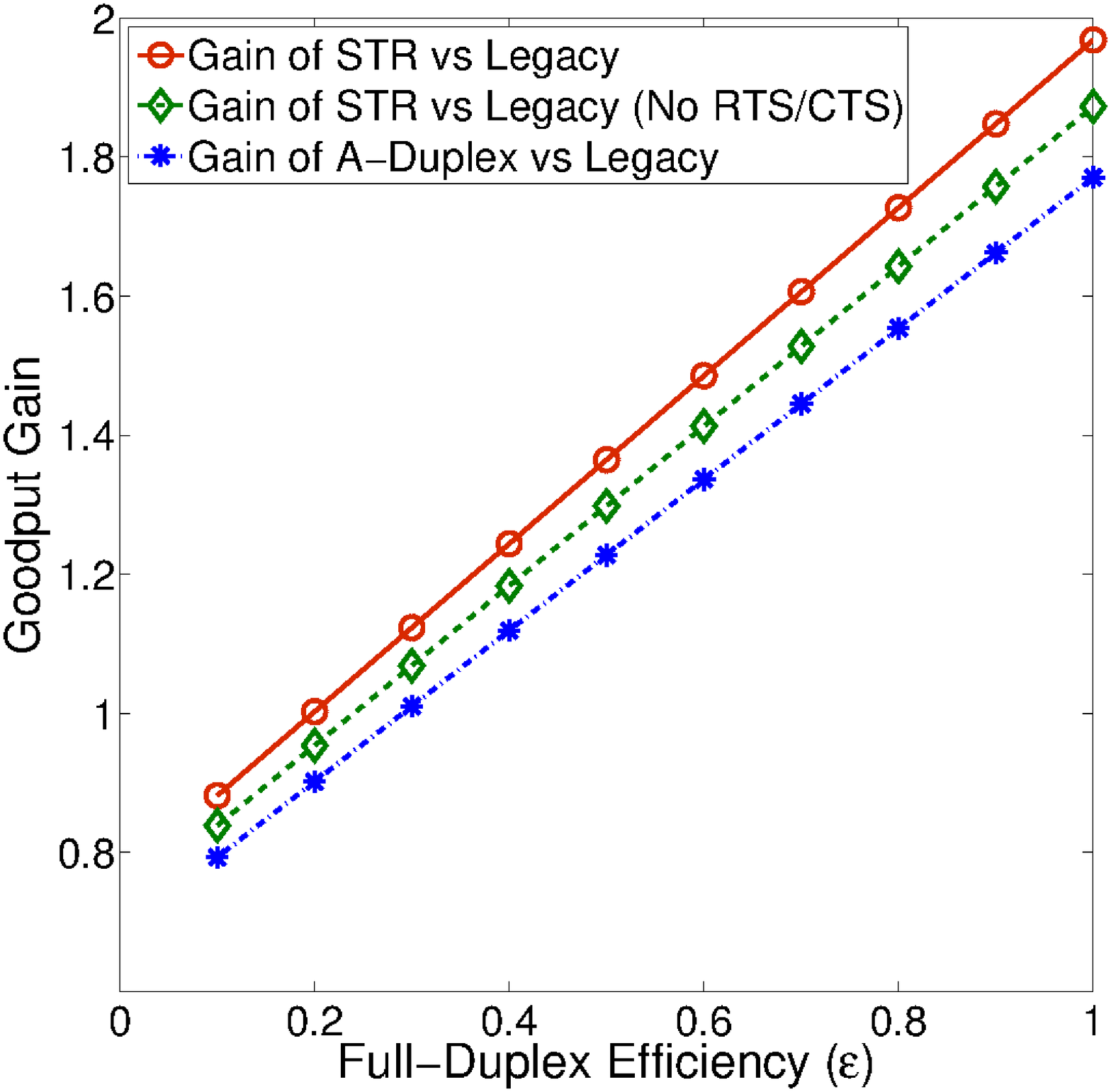}} \qquad
\subfloat[]{\label{cui_f}\includegraphics[scale=0.25]{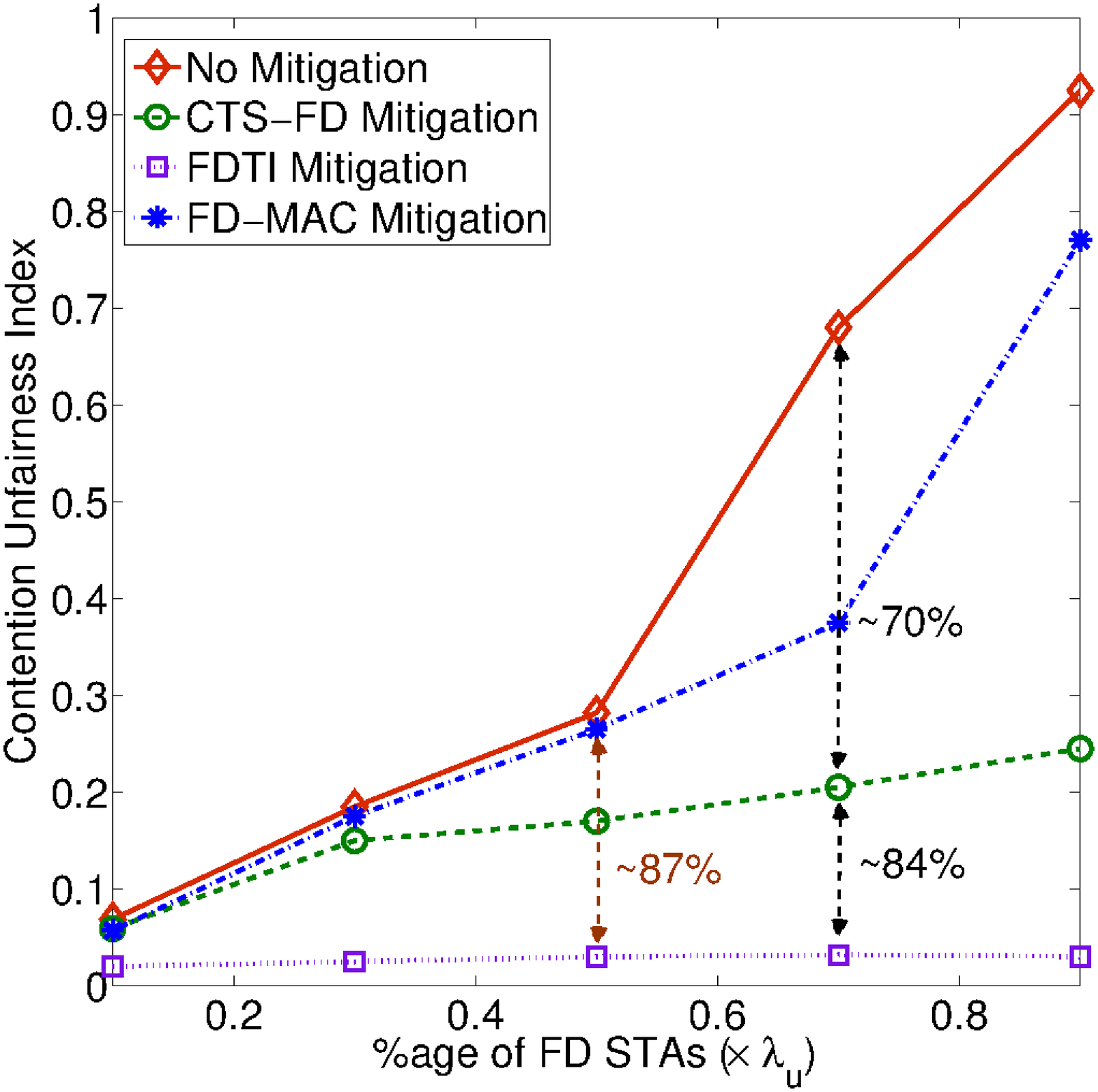}}
\caption{(a): STR Gain against threshold SINR for BFD case (\(\rho\) indicates self-interference cancellation capability and set to \(0.75\) unless otherwise specified); (b): STR Gain  against threshold SINR for UFD case; (c): Goodput gain of STR against FD efficiency (\(\text{SINR Threshold = 5}\ \text{dB}\)). Only UFD transmission is considered; (d): Contention unfairness index against density of FD STAs in case of BFD transmissions. }
\end{figure*}

We define the gain of enabling STR as \(\Theta_{STR}=\mathcal{T}_{STR}/\mathcal{T}_L\), where \(\mathcal{T}_{STR}\) and \(\mathcal{T}_L\) denote the average throughput of the network with STR and legacy mode of operation, respectively. Fig. \ref{STR_Gain} evaluates the gain of STR for the BFD case.  First, we note that the gain decreases as the threshold SINR increases. This is because the probability of a successful transmission, under the physical model, reduces with stringent SINR requirements. Second, we note that the STR gain increases initially as the density of FD-capable STAs, \(\lambda_f\), in the network increases. This is due to higher BFD transmissions in the network. However, as the SINR requirements become more stringent, the gain due to BFD transmissions reduces due to the fact that the probability of a successful BFD transmission decreases. Note that the STR gain with a lower \(\lambda_f\) is less sensitive to variation in SINR requirements as compared to the gain achieved under higher \(\lambda_f\). This is due to the fact that a FD transmission (either BFD or UFD) is successful when both first and second transmissions are successful. Note that a gain of nearly 2 is achieved when the density of FD-capable STAs is \(100\%\) in the network. Finally, we note that the gain is  dependent on the RSI. A lower self-interference cancellation capability, as indicated by \(\rho=0.6\), reduces the gain of STR due to higher RSI. 

Fig. \ref{STR_Gain_UFD} evaluates the gain of STR for the UFD case. This scenario is of particular interest when the density of FD-capable STAs in the network is low or negligible. The results, which follow a similar trend against threshold SINR as before, show that by enabling UFD transmissions in the network and thereby sharing the FD capability of the AP between STAs, a gain of nearly \(2\) is achievable, compared to legacy mode of operation. Note that the gain in this case depends on the carrier sensing range of a STA. The gain reduces as the carrier sensing range of a STA increases. This is due to the fact that a higher carrier sensing range reduces the probability of finding an eligible node for UFD transmission. One way of enhancing the gain of STR in UFD case is to use \emph{dynamic sensitivity control} (DSC) \cite{dsc_ref2} techniques, which are under active investigation within the 802.11ax working group. The fundamental principle of DSC is to dynamically tune the carrier sensing threshold  in order to improve spectral reuse. 

\begin{table*}
		\caption{Qualitative Comparison of Different MAC Protocols}
		\begin{center}
			\begin{tabular}{ccccc}
				\hline	
				\toprule
				\textbf{MAC Protocol} &  FD/HD Co-existence & Supports BFD & Supports UFD & Backwards Compatibility   \\\hline
				\midrule			
				\textbf{Janus} \cite{janus} & No & Yes & Yes & No \\
				\textbf{SRB-MAC} \cite{SRB_MAC} & No & Yes & Yes & No \\
				\textbf{FD-MAC} \cite{fd_mac} & Yes & Yes & No & No \\
				\textbf{A-Duplex} \cite{a-duplex} & No & No & Yes & No \\
				\textbf{PoC-MAC} \cite{PoCMAC} & No & No & Yes & No \\
				\textbf{CSR-MAC} \cite{csr_mac} & No & No & Yes & No \\
				\textbf{S-CW FD MAC} \cite{scw-fd} &  Yes & Yes & No & No\\
				\textbf{Proposed MAC}  & Yes & Yes & Yes & Yes \\
				\hline
			\end{tabular}
		\end{center}
			\label{qual_comp}
	\end{table*}

We evaluate the protocol overhead in terms of mean \emph{goodput} per transmission which can be considered as application-level throughput i.e., useful bits transmitted per unit of time. In order to capture the efficiency of the PHY layer self-interference cancellation technique at the MAC layer, we adopt the concept of \emph{FD efficiency} (\(\epsilon\))\cite{RTS_FCTS}, which is defined as the ratio of effective received packet payload to the sent packet payload. The  effective packet payload (in bits) for  BFD and UFD transmissions is given by  \(\mathcal{D}_{BFD}=\epsilon\left(\mathcal{D}_{FT}+\mathcal{D}_{ST}  \right)\) and \(\mathcal{D}_{UFD}=\epsilon\mathcal{D}_{FT}+\mathcal{D}_{ST}\) such that \(\mathcal{D}_{FT}\) and \(\mathcal{D}_{ST}\) denote the effective packet payload for the first transmission and second transmission, respectively.
The FD efficiency can be directly mapped to RSI model in \cite{exp_FD}. Fig. \ref{gpt_gain} evaluates the goodput gain of enabling STR mode. Note that the goodput gain is highly dependent on the performance of underlying PHY layer self-interference cancellation technique. A goodput gain of nearly 2 is achieved under perfect self-interference cancellation i.e., when \(\epsilon=1\). We also note that the goodput gain of STR is \(5\%\) lower compared to a legacy system employing no RTS/CTS for initiating transmissions. \textcolor{black}{Further, A-duplex \cite{a-duplex} provides up to  \(9\%\) lower goodput gain  as compared to the proposed MAC protocol. This is because a successful FD transmission in A-duplex must account for the additional time for capture effect. Besides, A-duplex employs a modified RTS frame with additional field.  }

Inspired by the Jain's fairness index, we propose \emph{contention unfairness index} (CUI) to measure contention unfairness in the network. 
The CUI is calculated in a similar way as Jain's fairness index; however it uses the fairness metric of \(f_i=w_i/\text{EIFS}\) such that \(w_i\) denotes the waiting time for the next contention of the \(i^{th}\) STA.
In Fig. \ref{cui_f}, we evaluate the CUI. We note that without any mitigation technique, contention unfairness could pose a severe challenge. The results demonstrate that while CTS-FD based contention unfairness mitigation provides a partial solution, FDTI-based contention unfairness mitigation is much more effective in resolving this issue. \textcolor{black}{Besides, FD-MAC \cite{fd_mac} mitigation technique, wherein the two nodes engaged in a FD transmission wait for EIFS after every successful transmission and other FD nodes do not ignore erroneous packets during NAV period, is also not effective, especially when the density of FD STAs in the network is high. }

Finally, a qualitative comparison of the proposed protocol against state-of-the-art is conduced in \tablename~\ref{qual_comp}.

\section{Concluding Remarks}\label{sect_cr}
This paper proposed a novel MAC protocol for enabling STR mode in 802.11 WLANs.  The proposed protocol provides a practical solution for achieving the benefits of FD technology in 802.11 WLANs while accounting for the co-existence of FD and legacy HD STAs, peculiarities of BFD and UFD links, and backwards compatibility with legacy protocols and STAs.  Performance evaluation reveals that significant performance gain can be achieved by enabling STR mode. The BFD transmissions achieve a gain of up to \(2\) depending upon the density of FD-capable STAs in the network. When the density of FD-capable STAs is low or the traffic is largely asymmetric, performance gain  is achievable by sharing the FD capability of the AP between HD or FD STAs through  UFD transmissions. Results also reveal that the proposed protocol incurs negligible overhead and provides a goodput gain of up to \(1.9\), depending on the efficiency of  self-interference cancellation techniques. The proposed solution is agnostic to the particular 802.11 variant  and provides a viable approach to achieve the benefits of FD technology in next generation WLANs.







\bibliographystyle{IEEEtran}

\bibliography{IEEEabrv,mybibfile}
%
\begin{IEEEbiographynophoto}{Adnan Aijaz} (M'14) received his Ph.D degree in Telecommunications Engineering from King's College London (KCL), UK, in  2014. After a post-doctoral year at KCL, he moved to Toshiba Research Europe Ltd. where he is currently  a Senior Research Engineer. His recent research interests include 802.11-based WLANs, 5G cellular networks, Industrial IoT, Tactile Internet, and full-duplex communications. His publications have been featured in internationally renowned conferences and journals.

Prior to joining KCL, he worked in cellular industry for nearly 2.5 years in the areas of network performance management, optimization, and quality assurance. He  holds a B.E. degree in Electrical (telecom) Engineering from National University of Sciences and Technology (NUST), Pakistan.
\end{IEEEbiographynophoto}

\begin{IEEEbiographynophoto}{Parag Kulkarni} is a Principal Research Engineer with Toshiba’s Telecommunications Research Laboratory in Bristol, UK, where he is actively engaged in work in the IoT / M2M and next generation wireless systems areas. He is broadly interested in technology and the role it can play to solve problems in different domains. During the course of his career so far, he has worked in a start-up and industry taking on responsibilities in the wide spectrum from ideation all the way through to technology transfer. He is a (co)inventor on over a dozen patents, co-recipient of Toshiba’s R\&D award for his work on wireless mesh networks for smart metering which led to a product and was also nominated to represent Toshiba Corporation at the inaugural edition of the Queen Elizabeth Prize for Engineering awards ceremony held at Buckingham Palace in 2013. He enjoys engaging in outreach activities targeted to both generalist and specialist audiences and, actively contributes to the research community in various capacities. Parag received his Bachelor of Technology (B.Tech) in Computer Science and Engineering from the National Institute of Technology Calicut India and a Ph.D from the University of Ulster in Northern Ireland.
\end{IEEEbiographynophoto}

\end{document}